\documentclass{article}
\usepackage{epsf}
\begin{document}

\title{ Can Low Mass Scalar Meson Nonet Survive in Large $N_c$ 
Limit ?}
\author{Masayuki \sc Uehara\footnote{E-mail: ueharam@cc.saga-u.ac.jp}\\
Takagise-Nishi 2-10-17, Saga 840-0921, Japan}
\date{\today}
\maketitle
\begin{abstract}
We study, within an approximate Inverse Amplitude Method to 
unitarize Chiral Perturbation Theory, whether low mass scalar mesons 
can survive in large $N_c$ regime, and show that vector mesons such as 
$\rho$ and $K^*$ survive as narrow width resonances, but all of the scalar 
meson nonet below 1GeV fade out as $N_c$ becomes large.
 \end{abstract}
 
\def\mib#1{\mbox{\boldmath{$#1$}}}
Recently Pel\'aez~\cite{Pelaez} has obtained an interesting result 
that the complex poles on the second Riemann sheet, corresponding 
to the  $\rho$ and $K^*$ mesons,  move towards the real axis as 
$N_c$ becomes large,  but in contrast those corresponding to the
 $\sigma$ and $\kappa$ states move away from the real axis. The study 
is performed within the Inverse Amplitude Method~(IAM) to unitarize 
one-loop Chiral Perturbation Theory~(ChPT) 
amplitudes~\cite{DP,Hanna,Guerr,GNP,PGN}. 
It is well known~\cite{tHooft,Witten} that a planar amplitude of 
$q\bar q$ meson-meson scattering is of $O(N_c^{-1})$ and that widths 
of intermediate $q\bar q$ mesons are also of $O(N_c^{-1})$. 
Thus, the result is consistent with the common understanding that 
the members of the vector meson nonet including  $\rho$ and $K^*$ are 
typical $q\bar q$ mesons. On the other hand the behavior of scalar 
mesons such as $\sigma$ and $\kappa$ is completely at variance with 
the nature expected for $q\bar q$ mesons. 

Inspired by Pel\'aez's observation we study how amplitudes
of two-NG boson scattering behave on the physical axis as $N_c$ 
increases within the approximate Oller-Oset-Pel\'aez~(OOP) version 
of IAM~\cite{OOP,MU04}, and how complex poles of $f_0(980)$ and 
$a_0(980)$ move in an approximate manner. In the OOP version only 
polynomial terms with the low energy constants $L_n$'s~(LEC's)  and 
$s$-channel loop terms  are taken into account out of full $O(p^4)$ 
amplitudes. 

In order to perform the study we have to find the explicit $N_c$ dependence 
in the scattering amplitudes.
Since the pion decay constant $f_\pi$ is of $O(N_c^{1/2})$ and the LEC's 
are to be of $O(N_c)$, except for $L_4$ and $L_6$, both of which are of 
$O(1)$\cite{GL,Espriu, Bijnens}, 
we fix the values of $L_n/f_\pi^2$ to those at $N_c=3$ and put 
$f_\pi(N_c)=\sqrt{N_c/3}\times f_\pi(3)$ with being $f_\pi(3)=93$ MeV.  
Indeed, the explicit large $N_c$ model calculations give the result that
 $L_n/f_\pi^2$ is $N_c$ independent for $n=1$ to 8, including 
 $L_7$~\cite{Espriu,Bijnens} .  $L_7$ is not of leading order, if 
 we consider 
 the $\eta'-\eta$ mixing~\cite{GL}, but we discard the $\eta'-\eta$ mixing 
 and regard $L_7$ as $O(N_c)$\cite{Espriu}, though we use the empirical 
 value for the $\eta$ mass.

The ingredients of IAM consist of the partial wave amplitudes with a definite 
isospin $I$ calculated  in terms of the amplitudes of ChPT of $O(p^2)$ and 
$O(p^4)$. The former amplitude $T^{(2)}(s,t,u)$ is written  
$O(p^2/f_\pi^2)$,  
and then of $O(N_c^{-1})$. The latter $O(p^4)$ amplitudes have polynomial 
terms with LEC's, which are written 
\begin{equation}
T^{(4)}_{\rm poly}(s,t,u)=\sum_{n=1,8}\frac{1}{f_\pi^2}
\left(\frac{L_n}{f_\pi^2}\right)P_n(s,t,u),
\end{equation}
where $P_n$'s are the second order polynomial terms of $s~t,{~\rm or}~u$ 
and meson mass squared. The polynomial term $T^{(4)}$ is of $O(1/N_c)$, 
because we fix  $L_n/f_\pi^2$'s to the values at $N_c=3$. 
The $s$-channel loop terms given as $t^{(2)}(s)J(s)t^{(2)}(s)$ are proportional 
to $O(s^2/f_\pi^4)$, where $J(s)$ is the one-loop function regularized 
as the 
$\overline{MS}-1$ scheme at the renormalization scale $\mu$~\cite{GL}, 
and $t^{(2)}$ is partial wave projected from $T^{(2)}(s,t,u)$. 
Since the $t$-  and  $u$-channel loop terms and tadpole terms are also of 
$O(p^4/f_\pi^4)$ , they are of $O(1/N_c^2)$. Thus, the OOP version is 
expected to be more valid as $N_c$ becomes large, because while the 
polynomial terms with LEC's are of $O(N_c^{-1})$, the discarded terms 
are of $O(N_c^{-2})$. The $s$-channel terms are crucial to realize 
unitarity, though they are  of $O(N_c^{-2})$.
This different $N_c$ dependence produces  different behavior of the 
amplitudes in the large $N_c$ regime. We note that 
IAM  gives a complex pole at a reasonable point in meson-meson scattering 
amplitudes, but this does not mean the IAM can give a complete prediction 
on the existence  or non-existence of a resonance, because the amplitude 
depends on the values of LEC's, which are determined so as for the 
amplitudes to reproduce extensive low energy data including  resonant 
behaviors or prominent structures. In this sense the study of the 
large $N_c$ behavior could provide a good laboratory to test the 
nature of hypothesized resonances. 
 
The values of LEC's, which reproduce phase shifts reasonably consistent with 
the experimental data, are summarized in Table I. These phenomenological 
values deviate somewhat from those of the large $N_c$ model calculations, 
but we use the above formulation of the $N_c$ dependence of the IAM 
\begin{table}[h]
\label{tab:Ln}
\begin{flushleft}
\begin{tabular}{|c|r|r|r|r|r|r|}\hline
&$ L_1$&$ L_2$&$ L_3$&$ L_5$&$ L_7$&$ L_8$ \\ \hline
Large $N_c$ & 0.79&1.58&$-3.17$& 0.43&$-0.42$& 0.46\\ \hline
IAM~I&0.56&1.21&$-2.79$&1.4&$-0.44$&0.78\\ \hline
IAM~III&0.60&1.22&$-3.02$&1.9&$-0.25$&0.84\\ \hline
Ours&0.55&0.98&$-2.95$&0.81&$-0.25$&0.49\\ \hline
ChPT&$0.52\pm 0.23$& $0.72\pm 0.24$&$-2.70\pm 0.99$
&$0.65\pm 0.12$&$-0.26\pm 0.15$& $0.47\pm 0.18$\\ \hline
\end{tabular}
\caption{$L_n\times 10^3$: The set Large $N_c$ is taken from 
Ref.~\cite{Espriu}, 
the sets IAM I and III for IAM with the full $O(p^4)$ amplitudes are taken 
from Ref.~\cite{Pelaez}, where $L_4=-0.36$ 
and $L_6=0.07$ in I and $L_4=0$ and $L_6=0.07$ in III, the set Ours is used 
in this work and  the set ChPT is the 2000 version of LEC's taken from 
Ref.~\cite{Amoros}. Used are $L_4=L_6=0$ in Large $N_c$, Ours and ChPT 
sets. The renormalization scale is $\mu=M_\rho$ except for Ours,which 
uses $\mu=1.07 $ GeV.}
\end{flushleft}
\end{table}
amplitudes. We point out that Our set is not chosen as the best solution to 
the overall fitting with the existing data. (Ours set in this work is 
slightly different from that of Ref.\cite{MU04}.)\\

\noindent {\bf Vector channels}\\
At first, we discuss the behaviors of vector mesons in the single channel 
formalism. The simple reason why the phase shift of the $\rho$ or $K^*$ 
channels increases across $\pi/2$ is due to the fact that the real part 
of the denominator, $t^{(2)}(s)-t^{(4)}_{\rm poly}(s)-t^{(2)}Re[J(s)]t^{(2)}$, 
develops a zero at the resonance position. We note that while the first  
and second terms behave as $O(1/f_\pi^2)$, the third term coming from 
the loop term does as $O(1/f_\pi^4)$. This implies that the zero almost
does not vary depending on the value of $N_c$. 
The zero depends dominantly on the combination of LEC's 
$2L_1-L_2+L_3$ as pointed out in Ref.~\cite{DP}. 
On the other hand the imaginary part $t^{(2)}\rho(s)t^{(2)}$ having a 
form $1/f_\pi^4$ becomes smaller and smaller as $N_c$ increases, 
where the phase space factor $\rho(s)={\rm Im}J(s)=k/(8\pi\sqrt{s})$ 
with $k$ being the CM momentum of two scattering mesons. 
\begin{center}
\begin{figure}[h!]
\epsfxsize=12 cm
\centerline{\epsfbox{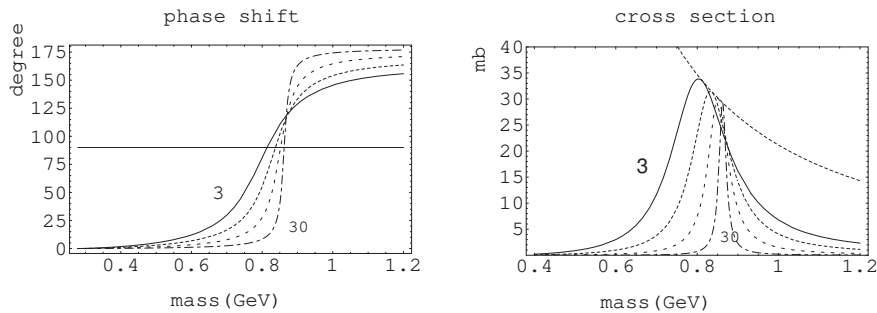}}
 \label{fig:rho}
\caption{$N_c$ dependence of phase shift (left) and 
 cross section (right) of the $\rho$ channel. Lines correspond to 
 $N_c=3$, 5, 10 and 30 from the top to the bottom. The inclined 
 doted line~(right) shows the kinematical limit of the cross section.}
\end{figure} 
\end{center}
Thus, the width decreases as like as $O(1/N_c)$, but the resonance 
position stays almost at the same point. 
This is the behavior on the physical real axis corresponding just to 
the behavior seen on the complex II sheet~\cite{Pelaez}. \\
\begin{center}
\begin{figure}[h!]
\epsfxsize=12 cm
\centerline{\epsfbox{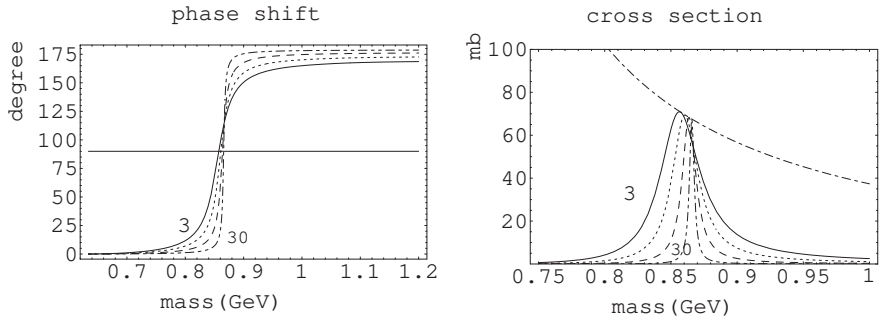}}
 \label{fig:rho}
\caption{$N_c$ dependence of phase shift (left) and 
 cross section (right) of the $K^*$ channel. Lines correspond to 
$N_c=3$, 5, 10 and 30 from the top to the bottom. The inclined 
doted line~(right) shows the kinematical limit of the cross section.}
\end{figure} 
\end{center}
If we consider the meson-meson scattering in the two channel formalism 
with the same LEC's, 
the combination of LEC's $2L_1-L_2+L_3$ develops the zero at almost the 
same point for $r_{11}$, $r_{12}$ and $r_{22}$, where 
$r_{ij}=t^{(2)}_{ij}-Re[t^{(4)}_{ij}]$, 
as pointed out in Ref.~\cite{MU04}. Our LEC's set gives unfortunately a 
little bit larger mass value for $\rho$, but smaller one for $K^*$. The 
octet component of the isoscalar vector meson 
$|V_8>=1/\sqrt{3}|\omega>+2/\sqrt{3}|\phi>$ 
has a mass below the $K\bar K$ threshold  both in our model and IAM 
with the full $T^{(4)}$ amplitudes\cite{GNP} as if it is a bound state with 
the decreasing residue as $O(1/N_c)$,  so that  if the pole shifts above the 
$K\bar K$ threshold, the shrinking width should occur.  

Thus, we can conclude that the vector mesons 
explained by IAM  have the nature consistent with the $q\bar q$  mesons. \\

\noindent{\bf Scalar channels}\\
Now, we proceed to the studies on the scalar channels. \\
\noindent$(\mib{I},~\mib{J})=(0,~0)$\\
This channel contains the enigmatic $\sigma$ and the $f_0(980)$ states.  
Using the two-channel IAM formalism with the $(\pi\pi)$ and $(K\bar K)$ 
channels, we calculate the $N_c$ dependence of the phase shift and the 
cross section as shown in Fig. 3.  We remember that the $f_0(980)$ state 
is generated as a bound state in the $K\bar K$ channel, where the kaon 
loop contribution is crucial~\cite{MU04}.
If $N_c$ increases, however, the real part of the $s$-channel loop term 
becomes small and the bound state pole shifts to the $K\bar K$ threshold 
and gets into the unphysical sheet through the cut starting with the 
$K\bar K$ threshold. In contrast to the vector channels the $N_c$ 
dependence of the phase shift and cross section shows drastic changes, 
therefore; the sharp rise of the phase shift near the $K\bar K$ threshold 
seen at $N_c=$3 and 4 disappears even at $N_c=5$, and shows a cusp-like 
behavior, and then the phase shift and the cross section shrink to the null 
structure after the cusp behavior disappears. \\

Where does the $f_0(980)$ pole go to ? We approximately calculate 
the pole position by expanding the amplitude in powers of $k_2$, the 
momentum of the $K\bar K$ channel, up to the first order, and observe 
that the pole moves into the upper~(lower) half plane of the  IV sheet 
from the lower~(upper) half plane of the II sheet, winding around the branch 
point at $K\bar K$ threshold,  and  goes away from the real axis as  
shown in the left side of Fig. 4.   Of course, positions at large $N_c$ 
are not so reliable owing to the rough approximation. No matter where 
the pole moves to, it holds valid that the physical trace of $f_0$ vanishes 
in the large $N_c$ limit.  \\
\begin{center}
\begin{figure}[h]
\epsfxsize=12 cm
\centerline{\epsfbox{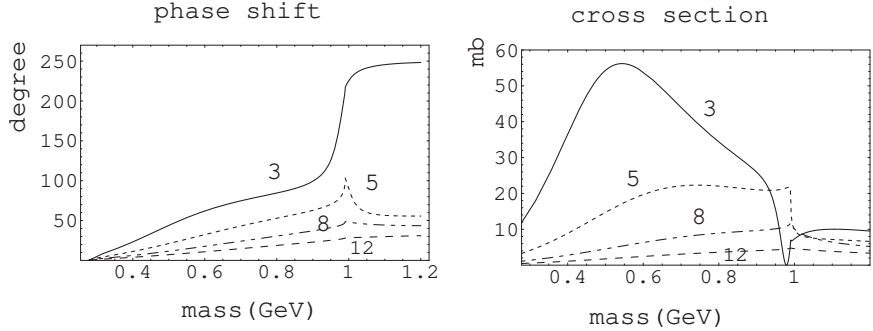}}
 \label{fig:fig00}
\caption{$N_c$ dependence of phase shift~(left) and 
 cross section~(right) of the (0,0) channel. Solid, dotted, dot-dot-dashed 
 and dashed lines are for $Nc=$3, 5, 8 and 12 respectively.}
\end{figure} 
\end{center}
\begin{center}
\begin{figure}[h!]
\epsfxsize=12 cm
\centerline{\epsfbox{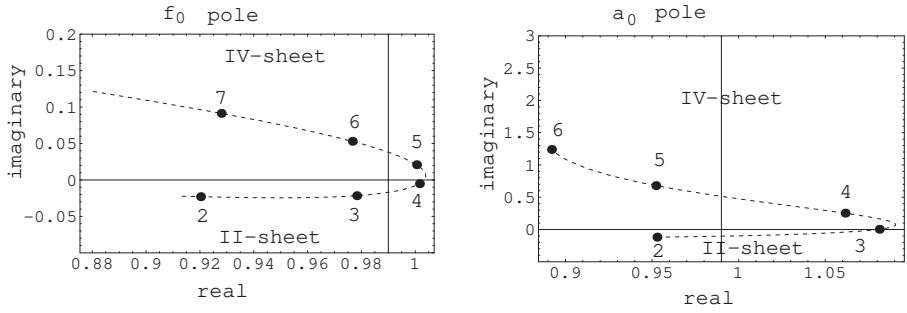}}
 \label{fig:pole}
\caption{$N_c$ dependence of the $f_0(980)$ pole~(left) and $a_0(980)$ 
pole~(right). Both of the poles wind aroundthe branch point at $K\bar K$ 
threshold to go to the upper half plane of the IV sheet. }
\end{figure} 
\end{center}
\begin{center}
\begin{figure}[h!]
\epsfxsize=12 cm
\centerline{\epsfbox{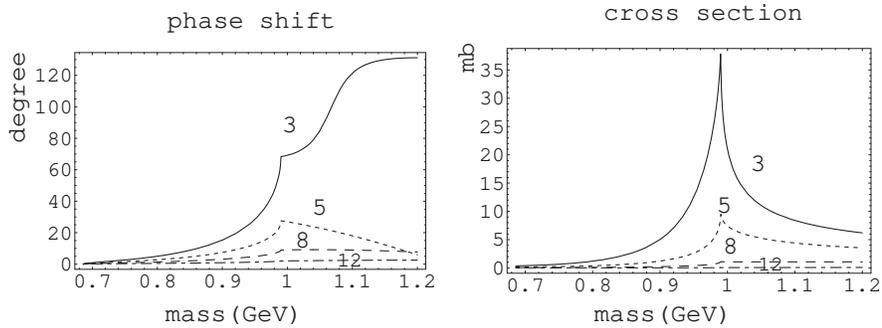}}
 \label{fig:fig(10)}
\caption{$N_c$ dependence of phase shift~(left) and 
 cross section~(right) of the (1,0) channel. $N_c=3$ , 4, 6 and 12 from 
 the top 
 to bottom. The vertical lines show the $K\bar K$ threshold.}
\end{figure} 
\end{center}
\noindent$(\mib{I},~\mib{J})=(1,~0)$\\
This channel contains the $a_0(980)$ state, which appears as a cusp-like 
sharp peak, because the complex pole sits above the $K\bar K$ threshold 
as $(1.071-i0.0198)$ GeV in the II sheet at $N_c=3$ in this work. 
This state  may be a typical example of the structure generated by t
he channel coupling  between the $\pi\eta$ and $K\bar K$ 
channel~\cite{MU04}.  As $N_c$ increases the rising phase shift 
above the cusp bends down  to a flat behavior and the cross section having  
a harp peak shrinks to the null structure. The pole moves from the sheet II 
to the sheet IV and leaves from the real axis as shown in the left side 
of Fig. 4.  The different pole movement between the isoscalar and 
isovector channel would come from the difference between the strong 
attractive $\pi\pi$ interaction in the former channel and the weak 
repulsive $\pi\eta$ one in the latter channel.\\

\noindent$(\mib{I},~\mib{J})=(1/2,~0)$\\
The null behaviors of the phase shift and cross section  for large $N_c$ 
are almost similar to the channels discussed above. 
The phase shift and the cross section are calculated in terms of the 
single $\pi K$ channel amplitude, because there appears a fictitious 
zero in the $\eta K$ amplitude in the OOP version. But we note that 
since the channel coupling between the $\pi K$ and $\eta K$ channel 
is weak, the results by the multichannel formalism are almost the 
same as those by the single channel formalism above the fictitious 
zero.  The calculations with use of the full $T^{(4)}$ give no such an 
unwanted zero~\cite{GNP} and the results are almost the same as ours. 
We emphasize that the $\pi K$  scattering cross section has a broad 
peak at 800 MeV at $N_c=3$  as well as the similar peak of the $\pi\pi$ 
cross section at 550 MeV. These peaks could be called as $\kappa$ and 
$\sigma$ ''resonances''. \\
\begin{center}
\begin{figure}[h!]
\epsfxsize=12 cm
\centerline{\epsfbox{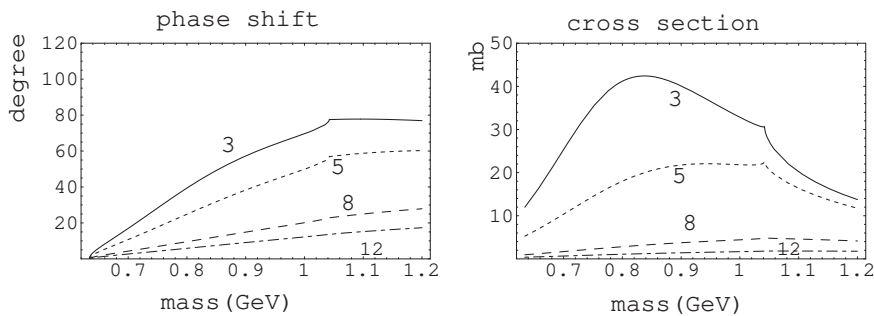}}
 \label{fig:fig(1/20)}
\caption{$N_c$ dependence of phase shift~(left) and 
 cross section~(right) of the (1/2,0) channel. $N_c=3$ , 5, 8 and 12 from 
 the above to bottom.}
\end{figure} 
\end{center}

We have calculated $N_c$ dependence of the vector and scalar meson 
channels though the approximate IAM under the rough  assumption that 
$L_n/f_\pi^2$'s are $N_c$ independent  and 
$f_\pi(N_c)=\sqrt{N_c/3}\times f_\pi(3)$. 
This assumption would be the more valid as $N_c$ becomes the 
larger, because the effects by remaining $O(1/Nc)$ terms would 
disappear, and then the results obtained here would remain valid i
n the full IAM calculations Thus, we observe  that all of the low mass 
scalar mesons including the $f_0(980)$ and $a_0(980)$ cannot survive 
in the large $N_c$ limit. This makes the sharp contrast to the vector 
mesons, which survive as the resonances with the extremely narrow 
widths. \\
 
Our conclusion is that the vector meson nonet has the nature consistent 
with the $q\bar q$ meson in large $N_c$ QCD, but the low mass scalar 
meson nonet cannot survive in the large $N_c$ limit and then does not 
have  the $q\bar q$ nature.  
Finally, we emphasize that  the scalar meson nonet should not be 
understood as  particles, which can  propagate with a definite mass 
and coupling constants, but as dynamical effects generated in coupled 
channel meson-meson scattering covering wide mass ranges  below 1 GeV, 
where chiral symmetry and unitarity play crucial roles. Their structures 
reveal themselves or not depending strongly on reactions, therefore. 
 Chiral unitary approach  provides a consistent formalism, which  
 realizes  such a picture in almost all of the scattering and production  
 processes.


\begin{thebibliography}{99}
\bibitem{Pelaez} J.R. Pel\'aeZ, arXiv:hep-ph/0307018; arXiv:hep-ph/0.06063.
\bibitem{DP}  A. Dobado and J.R. Pel\'aez, Phys. Rev. D56~(1997), 3057.
\bibitem{Hanna} T. Hanna, Phys. Rev. D 54~(1996), 4645; 
ibid. 55 (1997), 5613.
\bibitem{Guerr} F. Guerrero and J.A. Oller, 
Nucl. Phys. B~537 (1999), 459; Erratum; ibid. 602 (2000), 1641.
\bibitem{GNP} A. Gom\'ez Nicola and J.R. Pel\'aez,Phys. Rev. D~65 (2002), 
054009.
\bibitem{PGN} J.R. Pel\'aez and A. Gom\'ez Nicola, AIP Conf. Proc. 
{\bf 660} (2002), 102  arXiv:hep-ph/0.01049.
\bibitem{OOP} J.A. Oller, E. Oset and J.R. Pel\'aez, Phys. Rev. D~59 (1999), 
074001; Errata; ibid. 60~(1999), 099906; 62~(2000,)114017.
\bibitem{MU04} M. Uehara, arXiv:hep-ph/0204020.
\bibitem{tHooft}  G. t'Hooft, Nucl. Phys. B~72~(1974), 461; ibid. 
75~(1974), 461.
\bibitem{Witten} E. Witten, Nucl. Phys. B~160~(1979), 57.
\bibitem{GL} J. Gasser and H. Leutwyler, Nucl. Phys.B~250~)1985), 465.
\bibitem{Espriu} D. Espriu, E. de Rafael and J. Taron, Nucl. Phys. 
B~345~(1990), 22; Erratum; ibid. 355~(1991), 278. 
\bibitem{Bijnens} J. Bijnens, C. Bruno and E. de Rafael, Nucl. Phys. 
B~390~(1993), 501.
\bibitem{Amoros} G. Amor\'os, J. Bijnens,and P. Talavera, Nucl. Phys. 
B~602~(2001), 87.

\end{thebibliography}
\end{document}